\documentstyle[prl,aps,epsf,multicol,amssymb]{revtex}

\def\beginwide{
        \end{multicols} \vspace*{-0.5cm} \noindent
        \rule{3.5in}{.1mm}\rule{.1mm}{5mm} \widetext \medskip }
\def\beginwidetop{
        \end{multicols} \vspace*{-0.5cm} \noindent
        \widetext \medskip }
\def\endwide{
        \hspace*{3.5in}~\rule[-5mm]{.1mm}{5mm}\rule{3.5in}{.1mm}
        \begin{multicols}{2} \vspace*{-1.0cm} \noindent }
\def\endwidebottom{
        \begin{multicols}{2} \vspace*{-1.0cm} \noindent }

\begin{document}
\advance\textheight by 0.5in
\advance\topmargin by -0.25in
\draft

\title{Roughening Transition of Interfaces in Disordered Systems}

\author{Thorsten Emig$^a$ and Thomas Nattermann$^{a,b}$}
\address{$^a$Institut f\"ur Theoretische Physik, Universit\"at zu K\"oln,
  Z\"ulpicher Str. 77, D-50937 K\"oln, Germany\\
$^b$Ecole Normale Sup\'{e}rieure,
Laboratoire de Physique Th\'{e}orique,
24 rue Lhomond,
75231 Paris Cedex 05, France}

\date{\today}

\maketitle

\begin{abstract}
  The behavior of interfaces in the presence of both lattice pinning
  and random field (RF) or random bond (RB) disorder is studied using
  scaling arguments and functional renormalization techniques. For the
  first time we show that there is a continuous disorder driven
  roughening transition from a flat to a rough state for internal
  interface dimensions $2<D<4$.  The critical exponents are calculated
  in an $\epsilon$--expansion. At the transition the interface shows a
  superuniversal logarithmic roughness for both RF and RB systems. A
  transition does not exist at the upper critical dimension $D_c=4$.
  The transition is expected to be observable in systems with dipolar
  interactions by tuning the temperature.
\end{abstract}


\begin{multicols}{2}
  
  The thermal roughening transition (RT) from a flat, localized to a
  rough, delocalized interface in the three dimensional Ising and
  related lattice models is one of the paradigms of condensed matter
  physics \cite{Chaikin}.  It can be directly observed if a crystal
  surface undergoes a transition from a faceted to a smooth shape
  \cite{Weeks,Beijeren}. At the RT temperature $T_R$ the free energy
  of a step on the surface vanishes. Apart from its shape also other
  physical properties are influenced by the presence of the RT, e.g.,
  the (growth) velocity ${u}$ of the weakly driven interface (surface)
  changes dramatically at the RT from ${u} \sim \exp(-C/f)$ for
  $T<T_R$ to ${u}\sim f$ for $T>T_R$. Here $f$ denotes the driving
  force density, which is proportional to the magnetic field in the
  case of Ising magnets and to the difference $\Delta \mu$ of the
  chemical potentials of the crystal and its melt in the case of
  crystal growth, respectively.  The thermal RT was shown to be in the
  same universality class as the metal-insulator transition of a
  2D-Coulomb gas \cite{Chui.Weeks76}, the Kosterlitz-Thouless
  transition of a 2D XY-model \cite{Knops77} (both with an inverted
  temperature axis) and the phase transition of the 2D sine-Gordon
  model \cite{Jose76}. Interestingly, Kosterlitz \cite{Kosterlitz76}
  and Forgacs {\it et al.} \cite{Lipowsky} considered the thermal RT
  for $D\le 2$ interface dimensions and found a superuniversal
  logarithmic roughness {\it at} the transition.  Finally, the thermal
  RT disappears in $D=1$ dimensions: $1$-dimensional interfaces are
  rough at all finite T \cite{Fisher.Fisher}.
  
  {\it Disorder} is an even more efficient source of interface
  roughening and physically important as well \cite{Halpin.Zhang}.
  Interface roughening due to disorder was considered to determine the
  lower critical dimension of the random field Ising model
  \cite{VillainGrinstein,Binder} and the mobility of domain walls in
  disordered magnets \cite{Villain2,Joffe.Vinokur,Nattermann.92}. It
  was argued that in the presence of disorder interfaces of dimensions
  $D\leq 2$ are always rough, but undergo a roughening transition as a
  function of the disorder strength for $D > 2$ dimensions
  \cite{Nattermann84}.  Bouchaud and Georges \cite{BouchaudGeorges92}
  considered explicitly the competition between lattice and impurity
  pinning in $2\leq D\leq 4$ dimensions using a variational
  calculation.  Surprisingly they found {\it three} phases: a weakly
  disordered flat, a glassy rough (GR) and an intermediate flat phase
  (GF) with strong glassy behavior. The transition between the GR and
  the GF phase turned out to be first order. For $D=2$ only the glassy
  rough phase was expected to survive, but that shadows of the two
  other phases would still be seen in the short scale behavior.
  However, it should be kept in mind that the variational calculation
  is in general an uncontrolled approximation which is known to give
  spurious first order transitions \cite{Pokrovsky} and Flory-like
  exponents as indeed found in \cite{BouchaudGeorges92}.
  
  It is therefore the aim of the present paper to reconsider this
  problem by using a functional renormalization group calculation in
  $D=4-\epsilon$ interface dimensions. It turns out that the glassy
  flat phase found in \cite{BouchaudGeorges92} is replaced by a
  crossover region with logarithmic roughness around the RT.  The
  model we consider is that of \cite{BouchaudGeorges92}:
\begin{equation}
{\cal H}=\gamma\int d^Dx \left[\frac{1}{2}\left(\nabla z\right)^2+
V_L(z({\bf x}))+V_R(z({\bf x}),{\bf x})\right].
\label{eq:hamiltonian}
\end{equation}
Here $z({\bf x})$ denotes the displacement of the $D$ dimensional
interface from a flat reference configuration, and $\gamma$ is an
elastic stiffness constant. For simplicity, the lattice potential can
be considered to be given by $V_L(z)=-va^2\cos(2\pi z/a)$; higher
harmonics could be taken into account but turn out to be irrelevant.
Below, we will measure all transverse lengths in units of $a/2\pi$ and
therefore put $a=2\pi$ in the following.  The random pinning potential
$V_R(z({\bf x}),{\bf x})$ is assumed to be Gaussian distributed with a
zero average value $ \overline {V_R(z,{\bf x})}=0$ and
\begin{equation}
 \overline {V_R(z,{\bf x}) V_R(z',{\bf x'})}=R(z-z')\delta({\bf x-x'}).
\label{correlations}
\end{equation}
The form of $R(z)$ depends on the type of the disorder; in particular
we have to distinguish between random field (RF) and random bond (RB)
disorder.  For RF's, to begin with, $R(z)\sim -\Delta_0 |z|$ for large
$z$ where $\sqrt{\Delta_0}$ denotes the strength of the random field
\cite{VillainGrinstein}.  For RB's the bare form of $R(z)$ is a
smeared out $\delta$-function of width $\xi_0$ and strength
$\Delta_0\xi_0^2$.  For simplicity we assume that $\xi_0$ is of the
order $a$.

We begin our calculation with some elementary considerations. For
$\Delta_0=0$ a step of height $2\pi$ in the otherwise planar interface
of linear dimension $L$ costs an energy $E_{\rm step}\simeq \gamma
\sqrt{v}L^{D-1}$. The step itself has a width $w_{\rm step}\simeq
v^{-1/2}$. Here and below we omit all prefactors of order unity.
Switching on RF disorder, we consider the energy of a hump of total
height $z$ in the interface consisting of $z/2\pi$ such steps. This
gives for large enough lattice pinning $v$
\begin{equation}
E_{\rm total}/\gamma\simeq \sqrt{v}L^{D-1}z-\sqrt{\Delta_0 L^{D}z}.
\label{humpenergy}
\end{equation}
The second part represents the interaction energy with the RF
\cite{VillainGrinstein}.  Minimizing (\ref{humpenergy}) we get
$z\simeq (L/L_R)^{(2-D)}$ with $L_R\simeq (v/\Delta_0)^{1/(2-D)}$. For
$D<2$ and $L>L_R$ the interface is therefore rough even in the
presence of the lattice potential.  Since the total step free energy
vanishes now on scales $L\simeq L_R$, the system is described on
larger scales by an effective {\it elastic} Hamiltonian with a
stiffness constant $\gamma_{\rm eff}$. A rough estimate for
$\gamma_{\rm eff}$ follows from balancing the elastic and the bare
step free energy on the scale $L_R$, which gives $\gamma_{\rm
  eff}\simeq \gamma L_R/w_{\rm step}$ \cite{Nattermann84}.  On scales
$L\gg L_R$ the hump height scales as $z\simeq (L/L_R)^{\zeta}$ where
$\zeta=(4-D)/3$ \cite{VillainGrinstein}.  To be meaningful, $L_R$
should be larger than the bare stepwidth, $L_R>w_{\rm step}$, a
condition which can be rewritten as $v>v_c\simeq \Delta_0^{2/(4-D)}$.
$L_R$ can then be cast in the more conventional form of a
Larkin-length $L_R=(v/v_c)^{1/(2-D)}\Delta_0^{-1/(4-D)}$.  If
$v\rightarrow v_c+$, $\gamma_{\rm eff}$ changes into $\gamma$. For
$v<v_c$ we can neglect the lattice pinning term in
(\ref{eq:hamiltonian}) completely and $v$ has to be replaced by $v_c$
in the above formulas.

So far we have considered the case $D<2$. For $D=2$ it was shown by
Binder \cite{Binder,comment2} that $L_R$ becomes exponentially large,
$L_R\simeq \exp(Cv/\Delta_0)$, with $C$ a numerical factor, and
interfaces are again rough for $L>L_R$.  Thus, there is no flat phase
in $D\leq 2$ dimensions.

For $D>2$ we have to expect from the previous considerations that the
interface becomes rough only for $v<v_c$, i.e., that there is a RT as a
function of $v$ (or $\Delta_0$) at $v\simeq v_c$. In the rough phase
the mean square displacement of the interface can be estimated again
from the energy
\begin{equation}
\frac{E_{\rm total}}{\gamma L^{D-2}}\simeq {z^2}+
\sqrt{v}Lze^{-z^2/4}-\sqrt{\Delta_0 z}L^{(4-D)/2}.
\label{humpenergy2}
\end{equation}
Here we have included both the elastic and the step energy term to
describe the crossover from $v\ll v_c$ to $v\simeq v_c$. Moreover, we
have treated the lattice pinning term in a self consistent harmonic
approximation to account for a situation in which $z \gtrsim 2\pi$.
In the rough phase, the hump height asymptotically scales with the
roughness exponent $\zeta$.  However, there is an intermediate length
scale region in which the second term of (\ref{humpenergy2}) is larger
than the first term and the roughness increases only logarithmically,
$z^2 \simeq 2(D-2)\ln(L/L_R)$. Finally for $v\gg v_c$ the interface is
always flat. Similar arguments apply to RB systems but the asymptotic
roughness cannot be obtained from a Flory argument.  Below, we will
show by a detailed renormalization group (RG) calculation in
$D=4-\epsilon$ dimensions that the qualitative picture obtained so far
is correct and that the crossover from logarithmic to power law
roughness happens at length scales $L\simeq \xi_\| \sim
(v_c-v)^{-1/2\sqrt{\epsilon}}$.

As was first shown in \cite{Fisher86.Balents93}, the proper treatment
of interface roughening requires a functional RG approach due to the
occurrence of infinitely many relevant operators. A finite lattice
pinning potential $V_L(z)$ breaks the continuous translational symmetry
and thus a functional RG treatment of both pinning potentials
generates also contributions to the elastic stiffness.  Balents and
Kardar \cite{Balents.Kardar94} have developed a RG approach in terms
of an eigenfunction expansion of the interactions for a similar case.
For the periodic lattice potential, this expansion is simply a Fourier
expansion. This simplifies the analysis of the initial functional form
of the RG flow since higher harmonics $\cos mz$, $m>1$, are strongly
irrelevant at the RT with scaling dimension
$\lambda_m=2-(2+\epsilon/4)m^2$. Therefore only the lowest harmonic
$\cos z$ is taken into account.  We choose to keep $\gamma$ fixed by
rewriting the renormalization of $\gamma$ as an additional
renormalization of $T$, $R(z)$ and $v$. The resulting RG flow
equations, valid to lowest order in $\epsilon$, read
\begin{eqnarray}
\frac{dT}{dl}&=&\left[2-D-2\zeta-\frac{1}{4}v^2 \Delta 
\right] T\equiv -\theta T,\\
\frac{dR}{dl}&=&\left[4-D-4\zeta-\frac{1}{2}v^2 \Delta \right] R(z)
+\zeta zR'(z)\nonumber\\
&&+\frac{1}{2}[R''(z)]^2-R''(0) R''(z),\label{Rflow}\\
\frac{dv}{dl}&=&\left[ 2-\frac{1}{2}\Delta \right] v
+\frac{1}{8}\Delta v^3.\label{vflow}
\end{eqnarray}
Here $l=\ln(1/q)$ and $\Delta=-R''(0)e^{2\zeta l}$.  We have set the
ultraviolet cutoff to $1$ and made the substitution $K_D R(z)
\rightarrow R(z)$ with $K_D^{-1}=2^{D-1}\pi^{D/2}\Gamma(D/2)$. The
factor $\exp(2\zeta l)$ in the definition of $\Delta$ stems from the
rescaling of the periodicity of the lattice pinning potential. Here
$\exp(\zeta l)$ has to be read as the abbreviation for $\exp(\int
\zeta dl)$ with $l$ dependent $\zeta$.

First, we remark that for $D>2$ thermal fluctuations turn out to be
irrelevant such that $T^*=0$ at all fixed points of interest.  For
$v=0$ three locally stable fixed points of Eq.~(\ref{Rflow}) are
known, each with their own basin of attraction. These are the RF, the
RB and the charge density wave fixed point
\cite{Fisher86.Balents93,GiamarchiLeDoussal95}.  By inspection of Eq.
(\ref{Rflow}) it is clear that a finite $v$ does not change the
functional structure of the flow of $R(z)$ since $v^2\Delta$
approaches a finite value at the fixed point as we will see below.
Therefore the {\it functional form} of the fixed point $R^*(z)$ does
not depend on $v$ and is given by a dimensionless function $r(u)$ with
a characteristic scale of order unity.  It is determined by the fixed
point condition $\alpha r(u)+\beta u r'(u) + r''(u)(r''(u)/2+1)=0$
with $r''(0)=-1$, where $\alpha$ and $\beta$ are numerical
coefficients which differ for RF and RB disorder, respectively. They
can be determined from the conditions that $r(u)\sim |u|$ at large $u$
for RF's and that $r(u)$ decays exponentially for RB's. For RF's we
obtain $\alpha/\beta=-1$, which is sufficient to determine the
exponents. For RB disorder $r(0)>0$ and we can choose $r(0)=1$ leading
to $\alpha=1/2$ by evaluating the above fixed point condition at
$u=0$.  $\beta=0.6244$ is determined numerically from the condition
that the function $r(u)$ vanishes exponentially for $u\to \infty$ and
$r(0)=1$.

To obtain the critical behavior near the RT, we assume that $R(z)$ has
already approached its functional fixed point form, i.e., we make the
Ansatz
\begin{equation}
R(z)=\Delta(l)\xi^2(l)e^{-4\zeta l}r\left(z/\xi(l)e^{-\zeta l}\right),
\label{ansatz}
\end{equation}
which is justified on sufficiently large length scales. Inserting this
Ansatz into Eq.  (\ref{Rflow}) and using the fixed point condition for
$r(u)$ we obtain an equation of the form $Ar(u)+Bur'(u)=0$ where $A$
and $B$ depend on $\Delta$, $\xi$ and its derivatives with respect to
$l$. This equation is fulfilled for all $u$ if and only if $A=B=0$,
which is equivalent to the following flow equations for $\Delta$ and
$\xi$,
\begin{eqnarray}
  \frac{d\Delta}{dl}&=&\epsilon\Delta-\left[\frac{\alpha+2\beta}
        {\xi^2}+\frac{1}{2}v^2\right]\Delta^2,
  \label{flow_Delta}\\
  \frac{d\xi}{dl}&=&\beta\frac{\Delta}{\xi}.
  \label{flow_xi}
\end{eqnarray}

The rough phase of the interface is described by a stable fixed point
with $v^*=0$, $\Delta^*/\xi^{*2}=\epsilon/(\alpha+4\beta)$. In this
phase, RB and RF systems are characterized by a roughness exponent
which reads in our parameter approach
$\zeta=\epsilon/(4+\alpha/\beta)$. With the above values for $\alpha$
and $\beta$ the known results $\zeta^{RF}=\epsilon/3$ and
$\zeta^{RB}=0.2083\epsilon$ \cite{Fisher86.Balents93} are exactly
reproduced. For $D=4$ we find that the interface roughness grows only
{\it sub-logarithmically} with the system size $L$, $\overline{\langle
  z^2 \rangle} \sim \ln^\sigma(L/a)$ with $\sigma=2/3$ for RF's and
$\sigma=0.4166$ for RB's.

For {\it finite} $v$ there is a new fixed point with
$\Delta^*=4+\epsilon/2$, $v^*=\sqrt{\epsilon/2}$ and $\xi^*=\infty$.
Notice that the perturbative RG is still justified at this fixed point
since $v$ is small.  Linearizing around this fixed point we obtain in
lowest order in $\epsilon$ the eigenvalues $\lambda_{\pm}=\pm
2\sqrt{\epsilon}$. Since the fixed point has an unstable direction, it
has to be associated with the RT.  In the flat phase the size of
typical excursions of the interface from the preferred minimum of the
lattice potential defines the longitudinal correlation length
$\xi_\|\sim |v-v_c|^{-\nu_\|}$ with $\nu_\|=1/2\sqrt{\epsilon}$.
Interestingly $\nu_\|$ does not dependent on $\alpha$, $\beta$ and is
therefore universal for RF and RB disorder. The same expression for
$\nu_\|$ has been obtained previously for the {\it thermal} roughening
transition but with $\epsilon=2-D$ \cite{Lipowsky}.

The interface roughness $K({\bf x})=\overline{\langle [z({\bf
    x})-z({\bf 0})]^2 \rangle}$ can be obtained in terms of the
effective and non-rescaled parameters $\tilde\Delta=\Delta
e^{-\epsilon l}$, $\tilde v=v e^{-2l}$ using $l=\ln(1/q)$.
Sufficiently close to the transition where $\xi_\| > |{\bf x}| > L_R$,
the interface roughness increases logarithmically, $K({\bf
  x})\sim\ln(|{\bf x}|/L_R)$ on both sides of the RT, see Fig.~1(d).
Beyond $\xi_\|$ the roughness crosses over to the power law $K({\bf
  x})\sim (|{\bf x}|/\xi_\|)^{2\zeta}$ in the rough phase. On the flat
side of the transition $K({\bf x})$ saturates on scales larger than
$\xi_\|$ at a finite value $\sim \ln(\xi_\|/L_R)$.

The RG calculation also confirms the result
$v_c(\Delta_0)=A(\epsilon)\Delta_0^{2/\epsilon}$ already mentioned
above. To obtain the coefficient $A(\epsilon)$ we integrate the flow
equations (\ref{vflow}), (\ref{flow_Delta}) and (\ref{flow_xi}) by
neglecting the non-linear terms in $v$. The effect of the non-linear
terms is subsequently included by matching the resulting solution
$v(l)$ and $v^*$ at the length scale $L_R$, which yields
$A(\epsilon)=\frac{\epsilon}{2}(\frac{\epsilon}{\alpha+4\beta})^
{1/\zeta-2/\epsilon}(4e)^{-1/\zeta}$ in good agreement with numerical
solutions of the RG flow. Since $\zeta<\epsilon/2$, the function
$A(\epsilon)$ vanishes rapidly for $\epsilon \to 0$. Notice that in
$D=4$ an arbitrarily small $v>0$ leads to a flat phase due to the
sub-logarithmic roughness for $v\equiv 0$. The schematic RG flow for
$2<D<4$ is shown together with that expected for $D \le 2$ and $D\ge
4$ in Fig.~1.

The fluctuations in the free energy grows with length as $L^\theta$
where $\theta$ is the violation of hyperscaling exponent. Whereas one
has exactly $\theta=D-2+2\zeta$ in the rough phase, the relevance of
the lattice potential destroys the tilt symmetry of
(\ref{eq:hamiltonian}) leading to $\theta_c=2-\epsilon/2$ at the
transition.

\narrowtext
\begin{figure}[b]
\begin{center}
\leavevmode
\epsfxsize=1.0\linewidth
\epsfbox{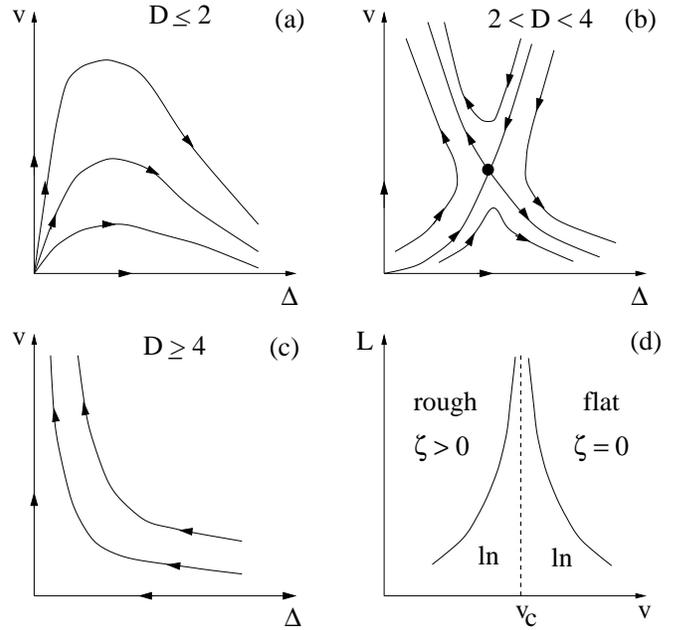}
\end{center}
\caption{Schematic RG flow as a function of the disorder variable
$\Delta$ and the lattice strength $v$ for (a) $D\le 2$, (b) $2<D<4$
and (c) $D\ge4$. Note that the fixed point associated with the rough
phase is shifted here to $\Delta=\infty$. (d) shows the length scale
$L$ dependent roughness near the RT.}
\label{fig.dia}
\end{figure}

As discussed in the introduction, the RT disappears in the physically
interesting case $D\to 2+$. However, the RT is expected to be seen
even for $D=2$ in systems with dipolar interaction. Indeed, for
magnetic or ferroelectric domain walls with the magnetization
direction $\rho$ parallel to the wall the elastic energy has the form
\cite{Lajzerowicz80,Nattermann83}
\begin{equation}
  \label{dipolar}
  E_{\rm el}=\frac{\gamma}{2}\int\frac{d^D{\bf q}}{(2\pi)^D}
  q^2\left(1+g\frac{q_\rho^2}{q^{D+1}}\right)
  z_{-{\bf q}}z_{\bf q},
\end{equation}
where $g$ measures the relative strength of the dipolar interaction.
Our results can also be applied to interfaces with this elastic
interaction. Repeating the RG analysis for this case, we see that the
calculated exponents remain valid to order $\epsilon$ if we replace
$\epsilon=4-D$ by $\epsilon=3(3-D)/2$. Thus the upper critical
dimension is shifted to $D=3$, and there is a RT for 2D dipolar
interfaces which is described by our results with $\epsilon=3/2$. Note
that 2D dipolar interfaces do not show a thermal RT
\cite{Lajzerowicz80}.

The threshold condition
$v_c(\Delta_0)\Delta_0^{-2/\epsilon}=A(\epsilon)$ and hence the RT can
be reached by changing the temperature $T$ since $v_0$ and
$\Delta_0^{2/\epsilon}$ depend in general in a different way on
$T_c-T$. Here $T_c$ denotes the condensation temperature of the
system, e.g., the Curie temperature for magnets. The precise
$T$-dependence is model dependent but close to $T_c$ where $\xi_0/a\gg
1$ we expect a very weak influence of the lattice potential, e.g., for
Peierls barriers $v_0\sim e^{-C\xi_0}$. Thus, for not too large
disorder, we can expect to see this RT by increasing $T$.

Under the influence of a small driving force density $f$, the
interface motion is dominated by jumps between neighboring metastable
states in the rough phase and between adjacent minima of the lattice
potential in the flat phase. In both cases the creep velocity $u$ is
exponentially small \cite{Joffe.Vinokur}, $u(f)\sim
\exp[-E_c/T(f_c/f)^\mu]$.  In the rough phase,
$\mu=(2\zeta+D-2)/(2-\zeta)$, $E_c\sim \gamma\xi_0^2 L_R^{D-2}$, and
$f_c\sim \gamma\xi_0 L_R^{-2}$ whereas in the flat phase $\mu=D-1$,
$E_c\sim \gamma v^{(2-D)/2}$ and $f_c\sim \gamma v$ for an interface
with short range interactions.  For dipolar systems, we obtain for a
two dimensional interface in the rough phase $\mu^{RF}=1$,
$\mu^{RB}=0.6666$ and for a flat interface $\mu=1$ independent of the
type of disorder. Therefore, in the RB case the phase of the interface
can be determined by measuring $\mu$. A very recent experiment on
driven domain walls \cite{Lemerle98} shows that the exponent $\mu$ can
be measured very accurately. 

To summarize, the interplay between a random potential and lattice
pinning for an interface leads for $2<D<4$ to a {\it continuous}
disorder-driven roughening transition between a rough and a flat
interface, in contrast to earlier results of Bouchaud and Georges for
RB systems \cite{BouchaudGeorges92} which are based on a variational
calculation. However, there is a qualitative similarity in the phase
diagram if we identify the GF phase of \cite{BouchaudGeorges92} with
the logarithmically rough crossover region of our results. Both
separate the GR from the simple flat phase \cite{comment1}. Moreover,
we have obtained the exponent $\nu_\|$ which has to lowest order in
$\epsilon$ the same form as that of the thermal RT if $\epsilon=2-D$.
For both random field and random bond systems, the interface shows a
superuniversal logarithmic roughness {\it at} the transition.  At the
upper critical dimension $D=4$ an arbitrary small lattice pinning
produces a flat interface due to sub-logarithmic roughness in the
absence of lattice effects. In contrast to the 2D thermal RT a
diverging correlation length $\xi_\|$ appears on both sides of the
transition. On the rough side $\xi_\|$ sets the length scale for the
crossover from logarithmic to algebraic roughness.  We have
demonstrated that the disorder-driven RT will appear for two
dimensional interfaces in dipolar systems.

We acknowledge discussions with A.~Georges and S.~Scheidl.  This
research was supported by the German-Israeli Foundation (GIF) and the
Volkswagen Stiftung.

\end{multicols}{}

\begin{references}                                
 
\bibitem{Chaikin} P.M.~Chaikin and T.C.~Lubensky, Principles of
  condensed matter physics (Cambridge UP, 1995).

\bibitem{Weeks} J.D. Weeks in {\it Ordering in Strongly Fluctuating Systems} 
ed. by T. Riste (Plenum, 1980).

\bibitem{Beijeren} H.~van Beijeren and I.~Nolden in {\it Structure and
    Dynamics of Surfaces II}, ed. by W.~Schrommers and P.~von
  Blanckenhagen (Springer, Berlin, 1987).

\bibitem{Chui.Weeks76} S.T.~Chui and J.D.~Weeks, Phys.~Rev.~B {\bf 14}, 
4978 (1976).

\bibitem{Knops77} H.J.F.~Knops, Phys.~Rev.~Lett.~{\bf 39}, 766 (1977). 

\bibitem{Jose76} J.~Jose, Phys.~Rev.~D~{\bf 14}, 2826 (1976).

\bibitem{Kosterlitz76} J.M.~Kosterlitz, J.~Phys.~C~{\bf 10}, 3753 (1977).
  
\bibitem{Lipowsky} G.~Forgacs, R.~Lipowsky and T.M.~Nieuwenhuizen, in
  {\it Phase Transitions and Critical Phenomena}, Vol.~14, ed. by
  C.~Domb and J.L.~Lebowitz (AP, London, 1991).

\bibitem{Fisher.Fisher} D.S.~Fisher and M.E.~Fisher, Phys.~Rev.~B~{\bf 25},
3192 (1982).

\bibitem{Halpin.Zhang} For a review see T.~Halpin-Healy and Y.C.~Zhang,
Phys.~Rep.~{\bf 254}, 215 (1995).

\bibitem{VillainGrinstein} J.~Villain, J.~Phys.~(Paris) {\bf 43}, L551
  (1982); G.~Grinstein and S.K.~Ma, Phys.~Rev.~Lett.~{\bf 49}, 685
  (1982).

\bibitem{Binder} K.~Binder, Z.~Phys.~B {\bf 50}, 343 (1983).
  
\bibitem{Villain2} J.~Villain, Phys.~Rev.~Lett.~{\bf 52}, 1543 (1984).

\bibitem{Joffe.Vinokur} L.B.~Ioffe and V.M.~Vinokur, J.~Phys.~C~{\bf 20}, 
6149 (1987).

\bibitem{Nattermann.92} T.~Nattermann et al., J.~Phys.~II~France~{\bf 2}, 1483 
(1992).

\bibitem{Nattermann84} T.~Nattermann, Z.~Phys.~B {\bf 54}, 247 (1984); 
Phys.~Stat.~Sol.~(b)~{\bf 132}, 125 (1985).

\bibitem{BouchaudGeorges92} J.P.~Bouchaud and A.~Georges, Phys.~Rev.~Lett.~
{\bf 68}, 3908 (1992).

\bibitem{Pokrovsky} see e.g., A.~Patashinsky and V.~Pokrovsky, {\it
    Fluctuation Theory of Phase Transitions} (Pergamon Press, 1979).
  
\bibitem{comment2} Binder considers actually the surface of an Imry-Ma domain 
in the 2D RF Ising model which corresponds in our case to the boundary 
of an elevated terrace on the 2D interface. 

\bibitem{Fisher86.Balents93} D.S.~Fisher, Phys.~Rev.~Lett. {\bf 56},
  1964 (1986); L.~Balents and D.S.~Fisher, Phys.~Rev.~B {\bf 48}, 5949
  (1993).

\bibitem{Balents.Kardar94} L.~Balents and M.~Kardar, Phys.~Rev.~B {\bf
    49}, 13030 (1994).

\bibitem{GiamarchiLeDoussal95} T.~Giamarchi and P.~Le Doussal,
  Phys.~Rev.~B {\bf 52}, 1242 (1995).

\bibitem{Lajzerowicz80} J.~Lajzerowicz, Ferroelectrics {\bf 24}, 179 (1980).

\bibitem{Nattermann83} T.~Nattermann, J.~Phys.~C~{\bf 16}, 4125 (1983).

\bibitem{Lemerle98} S.~Lemerle {\it et al.}, Phys.~Rev.~Lett.~{\bf
    80}, 849 (1998).

\bibitem{comment1} The generalization of our results to a $N$
  component displacement yields also a continuous transition as long
  as $N<\infty$ and a first order transition for $N=\infty$ where the
  variational approach is expected to become exact.

\end{references}
\end{document}